\documentclass[a4paper,12pt]{article}
\usepackage[utf8]{inputenc}
\usepackage{geometry}
\usepackage[sumlimits,intlimits,namelimits]{amsmath}
\usepackage{amssymb}
\usepackage{a4wide}
\usepackage[english]{babel}
\usepackage{longtable, multirow,tabularx}
\usepackage[titletoc,title]{appendix}
\usepackage{slashed}
\usepackage{amsmath}
\usepackage{cite}

\makeatletter
\def\fmslash{\@ifnextchar[{\fmsl@sh}{\fmsl@sh[0mu]}}
\def\fmsl@sh[#1]#2{%
  \mathchoice
    {\@fmsl@sh\displaystyle{#1}{#2}}%
    {\@fmsl@sh\textstyle{#1}{#2}}%
    {\@fmsl@sh\scriptstyle{#1}{#2}}%
    {\@fmsl@sh\scriptscriptstyle{#1}{#2}}}
\def\@fmsl@sh#1#2#3{\m@th\ooalign{$\hfil#1\mkern#2/\hfil$\crcr$#1#3$}}
\makeatother
\usepackage{graphicx}
\usepackage{subfigure}
\usepackage{epstopdf}
\usepackage{xcolor}

\begin{document}
\begin{titlepage}
\begin{flushright}
SI-HEP-2020-24 \\[0.2cm]
SFB-257-P3H-20-049
\end{flushright}

\vspace{1.2cm}
\begin{center}
{\Large\bf
Completing $1/m_b^3$ corrections to non-leptonic bottom-to-up-quark decays
}
\end{center}

\vspace{0.5cm}
\begin{center}
{\sc Daniel Moreno} \\[2mm]
{Theoretische Physik 1, Naturwiss. techn. Fakult\"at,
Universit\"at Siegen\\ D-57068 Siegen, Germany}  
\end{center}

\vspace{0.8cm}
\begin{abstract}
\vspace{0.2cm}\noindent
We compute the contributions of dimension six two-quark operators to the non-leptonic decay width of heavy hadrons due to the flavor changing 
bottom-to-up-quark transition in the heavy quark expansion. Analytical expressions for the Darwin term $\rho_D$ and the 
spin-orbit term $\rho_{\rm LS}$ are obtained with leading order accuracy.
\end{abstract}

\end{titlepage}

\newpage
\pagenumbering{arabic}
\section{Introduction}

The HQE~\cite{MW:00,G:04,N:94} provides a well-established theoretical tool to study hadrons containing a single heavy quark $Q$. 
In particular, it has been used intensively over the last decades to study inclusive decays of heavy 
hadrons~\cite{Lenz,Cheng:2018rkz,Krinner:2013cja}. The study of these systems is specially interesting because it requires understanding
the interplay between electroweak and strong interactions.
Within this framework, the decay width of the heavy hadron is written as a combined expansion in the strong coupling constant $\alpha_s (m_Q)$~\cite{BG:95} 
and the inverse heavy quark mass $m_Q$~\cite{EH:90}. The leading term of that expansion predicts that the decay of the heavy hadron is 
given by the weak decay of the "free" heavy quark it is made of~\cite{Bagan:1994zd,Beneke:1998sy,Beneke:2002rj}, whereas it is blind to the spectator 
quark it contains. As a consequence, lifetimes are predicted to be the same for all hadrons which contain the same heavy quark. 
The first correction to this picture appears at $\mathcal{O}(1/m_Q^2)$, which introduces lifetime differences between hadrons with different 
spin configurations. Lifetime differences caused by a different spectator quark flavour content
appear for the first time at $\mathcal{O}(1/m_Q^3)$ through four-quark operators, 
which explicitly contain spectator quarks of a particular flavour. 
These contributions are suppressed by the heavy quark mass and, as a consequence, they are expected to be small corrections to the "free" heavy quark decay.

Thus, the HQE provides a systematic way of improving theoretical predictions by either including matrix elements (ME) of 
higher order operators (including more terms in the $1/m_Q$ expansion)~\cite{Heinonen:2014dxa}
or improving the precision in $\alpha_s(m_Q)$ of the perturbative coefficients in front of them. 

When it comes to test it, the HQE has proven to be extremely successful to describe lifetimes and lifetime differences of 
bottomed hadrons, whose experimental determination has reached an impressive precision. The current (2019) experimental averages 
obtained by the Heavy Flavor Averaging Group (HFLAV) are~\cite{Amhis:HFLAV} 
\begin{equation*}
 \frac{\tau(B_s)}{\tau(B_d)}\bigg|^{\mbox{\scriptsize exp}} = 0.994 \pm 0.004\,,
 \;\;\;\; 
 \frac{\tau(B^+)}{\tau(B_d)}\bigg|^{\mbox{\scriptsize exp}} = 1.076 \pm 0.004\,,
 \;\;\;\; 
 \frac{\tau(\Lambda_b)}{\tau(B_d)}\bigg|^{\mbox{\scriptsize exp}} = 0.969 \pm 0.006\,,
\end{equation*}
and even higher precision seems to be achievable from the most recent results from LHCb~\cite{lhcb} and ATLAS~\cite{atlas}, 
so $B$-physics is in its precision era.

That is an incredible opportunity to test the HQE, especially for lifetime differences. 
Therefore, the theory precision should reach a similar status. At present,
the theoretical predictions are really good as well~\cite{Neubert:1996we,Lenz:2014,Lenz:2017}
\begin{equation*}
 \frac{\tau(B_s)}{\tau(B_d)}\bigg|^{\mbox{\scriptsize th}} = 1.0006 \pm 0.0025\,,
 \;\;\;\; 
 \frac{\tau(B^+)}{\tau(B_d)}\bigg|^{\mbox{\scriptsize th}} = 1.082^{+0.022}_{-0.026}\,,
 \;\;\;\; 
 \frac{\tau(\Lambda_b)}{\tau(B_d)}\bigg|^{\mbox{\scriptsize th}} 
= 0.935 \pm 0.054\,,
\end{equation*}
which is due to the following theoretical advancements. The leading order ("free" heavy quark decay) term is currently known at 
NLO-QCD~\cite{Hokim:84,Altarelli:91,Voloshin:95,Bagan:95,LenzNierste:97,LenzNierste:99,Krinner:2013cja,Bagan:1994zd} and at NNLO-QCD
in the massless case~\cite{Czarnecki:06} for non-leptonic decays. For semi-leptonic decays the current precision is 
NNLO-QCD~\cite{CzaneckiMelnikov:97,CzaneckiMelnikov:99,vanRitbergen:99,Melnikov:08,PakCzarmnecki:08,PakCzarmnecki2:08,Dowling:08,Bonciani:08,BiswasMelnikov:10,Caola:13}.

The contribution from the first power correction (dimension five operators) is 
known at NLO-QCD for semi-leptonic decays, whereas it is only known at LO-QCD for 
non-leptonic decays~\cite{Uraltsev:92,BlokShifman:93,BlokShifman2:93,BlokShifman:92,Gambino:14,Mannel:2014xza,Mannel:2015jka}.

The contribution from the second power correction (dimension six two-quark operators, also called Darwin 
and spin-orbit terms) is known at NLO-QCD~\cite{GremKapustin:97,Mannel:2017jfk,Mannel:2019qel} 
for semi-leptonic decays and at LO-QCD for non-leptonic decays\cite{Mannel:2020fts,Lenz:2020oce}.

Finally, the contribution from dimension six four-quark operators, which induces lifetime differences and which is enhanced 
by a phase space factor $16\pi^2$, is known at NLO-QCD~\cite{Beneke:2002rj,Mescia:2002,Lenz:2013}.
 
In this work, we compute the coefficients of the dimension six two-quark operators (Darwin 
and spin-orbit terms) for the Cabibbo suppressed channels of $\mathcal{O}(\lambda^3)$ 
stemming from $b\rightarrow u$ transitions. These coefficients were already computed in Ref. \cite{Lenz:2020oce} but not 
in Ref. \cite{Mannel:2020fts}. 
Our aim is to perform a check of the results presented there while 
using a completely different approach following the lines of Ref. \cite{Mannel:2020fts}.
The coefficients presented here may be relevant to improve the theoretical precision of $B$-hadron lifetimes and 
lifetime differences.

The results obtained here, especially for the $b\rightarrow u\bar u d$ channel, can be easily applied to charm decays, where their contribution 
is expected to be more important because the HQE for charm has a slower convergence than for bottom.
These coefficients might be helpful to clarify the status of the HQE for charm~\cite{Lenz:2013,Fael:2019umf}, whose validity has been often questioned due to 
the smallness of the charm quark mass.

The paper is organized as follows. In Sec. \ref{Sec:outline} we present the outline of the calculation. We give definitions in 
Sec. \ref{subsec:def} and compute the matching coefficients of the 
dimension six two- and four-quark operators at tree level in Secs. \ref{Sec:match2F} and \ref{Sec:match4F}, respectively.
Renormalization of the coefficients is discussed in Sec. \ref{Sec:renormalization}. 
In Sec. \ref{sect:results} we present the results and discuss the role of evanescent operators in Sec. \ref{Sec:eva}. Finally, we perform a 
numerical analysis in Sec. \ref{Sec:numerical}. We also give some technical results in the Appendix.

\section{Outline of the Calculation}
\label{Sec:outline}

\subsection{Definitions}
\label{subsec:def}

Weak decays of hadrons are mediated by the weak interactions of their quark constituents. These flavor changing transitions 
are described very efficiently through an operator product expansion (OPE) approach which gives rise 
to an effective Lagrangian~\cite{BBL:96}. Despite of the plethora of 
effective operators necessary to describe the weak interactions we only consider here the tree-level four-quark operators of 
current-current type, since their Wilson coefficients are the largest. The effective Lagrangian describing 
$b\rightarrow q_3 \bar q_1 q_2$ transitions reads

\begin{equation}
\label{Leff}
 \mathcal{L}_{{\scriptsize\mbox{eff}}} = - \frac{4G_F}{\sqrt{2}}V_{q_1 q_2} V_{q_3 b}^* (C_1 \mathcal{O}_1 + C_2 \mathcal{O}_2) 
 + \mbox{h.c}\,,
\end{equation}
where $G_F$ is the Fermi constant, ${\cal O}_{1}$, $\mathcal{O}_2$ are four-quark operators of the current-current type,
and $V_{q_1 q_2}, V_{q_3 b}^*$ are the corresponding Cabibbo-Kobayashi-Maskawa (CKM) matrix elements which describe 
the mixing of quark generations under weak interactions. However, not all the operators of this kind have the same importance. 
Their weight is determined by the size of the corresponding CKM matrix elements
\begin{eqnarray} \label{eq:Leff1}
 \mathcal{L}_{\mbox{\scriptsize eff}} &\sim&
 \lambda^2( {\bar b}c {\bar u} d + {\bar b} c {\bar c} s)
 \\
&&
+\lambda^3({\bar b} u {\bar u} d + {\bar b} u {\bar c} s + {\bar b}c {\bar u} s + {\bar b} c {\bar c} d )
\\
&&
+ \lambda^4({\bar b} u {\bar u} s + {\bar b} u {\bar c} d)\,,
\end{eqnarray}
where $\lambda=V_{us}=0.2257^{+0.0009}_{-0.0010}$ is a Wolfenstein expansion parameter. The contribution to the total non-leptonic width coming 
from the dominant $\mathcal{O}(\lambda^2)$ operators was already considered in Refs. \cite{Mannel:2020fts,Lenz:2020oce}. 
In this work, we consider the Cabibbo suppressed contribution of $\mathcal{O}(\lambda^3)$ stemming from the $b\rightarrow u$ transitions ($q_3=u$), 
which was already considered in Ref. \cite{Lenz:2020oce}. Our aim is to perform a check of the results presented there while 
using a completely different approach following the lines of Ref. \cite{Mannel:2020fts}. 
It is worth mentioning that the remaining Cabibbo suppressed contributions of $\mathcal{O}(\lambda^3)$ and $\mathcal{O}(\lambda^4)$ can be directly obtained 
from the results presented here and in Refs. \cite{Lenz:2020oce,Mannel:2020fts} by exchanging the $d$ and $s$ quarks\footnote{This statement is 
true if the light quarks $d$, $u$ and $s$ are considered to be massless, which is the case in the present work.}.

In summary, we are interested in weak decays of $B$-hadrons mediated by transitions of the type $b\to u\bar q_1 q_2$  
with $(q_1,q_2) = (u,d)$ and $(q_1,q_2) = (c,s)$. In that case, the operator basis reads~\cite{BBL:96}
\begin{equation}
\label{eq:canbasis}
 \mathcal{O}_1 = (\bar u^i \Gamma_\mu b^i)(\bar q_2^j \Gamma^\mu q_1^j)\,,\quad
 \mathcal{O}_2 = (\bar u^i \Gamma_\mu b^j)(\bar q_2^j \Gamma^\mu q_1^i)\,,
\end{equation}
with $\Gamma_\mu=\gamma_\mu (1-\gamma_5)/2$. However, for the purpose of the computation it is convenient to use an operator basis diagonal in the 
color space~\cite{Beneke:2002rj}
\begin{equation}
\label{eq:newbasis}
 \mathcal{O}_1 = (\bar u^i \Gamma_\mu b^i)(\bar q_2^j \Gamma^\mu q_1^j)\,,\quad
 \mathcal{O}_2 = (\bar q_2^i \Gamma_\mu b^i)(\bar u^j \Gamma^\mu q_1^j) \,,
\end{equation}
which is obtained after applying a Fierz tranformation in the operator $\mathcal{O}_2$. The computation is 
carried out using the standard techniques of dimensional regularization ($D= 4-2\epsilon$)~\cite{tHooft:1972tcz} 
with anticommuting $\gamma_5$~\cite{Chetyrkin:1997gb,Grozin:2017uto} and renormalization. Therefore 
the Dirac algebra of $\gamma$-matrices usually defined in $D=4$ needs to be extended to 
$D$-dimensional spacetime~\cite{Altarelli:1980fi,Buras:1989xd,Buras:1990fn,Chanowitz:1979zu}. 
A consequence of it is that using Fierz transformations can lead to non-trivial $\epsilon$ 
dependences~\cite{Dugan:1990df,Herrlich:1994kh,Pivovarov:1988gt,Pivovarov:1991nk} 
and in general a change in the operator basis valid in four-dimensional spacetime is not allowed in $D$-dimensional 
spacetime\footnote{Indeed we can enforce these tranformations to be valid in $D$-dimensional spacetime as well, but the price we have to pay is the 
appearance of additional operators called evanescent operators~\cite{Grozin:2017uto,Beneke:2002rj}. This statement will be discussed later in more detail.}. 
In our particular case, the transformation of Eq. (\ref{eq:canbasis}) into Eq. (\ref{eq:newbasis}) is allowed because at the order we are working in, 
only tree level expressions for the Wilson coefficients $C_1$ and $C_2$ are needed.

According to the optical theorem the $B$-hadron decay rate for the inclusive non-leptonic decays 
can be related to the discontinuity of the forward scattering matrix element
\begin{eqnarray}
 \Gamma(b\rightarrow u \bar q_1 q_2) &=& 
 \frac{1}{2M_B}\mbox{Im}\, \langle B(p_B)\lvert i \int d^4 x\, 
T\{\mathcal{L}_{\mbox{\scriptsize eff}}(x),\mathcal{L}_{{\scriptsize\rm eff}}(0)\} \lvert B(p_B)\rangle
\nonumber
\\
 &=& \frac{1}{2M_B}\langle B(p_B)\lvert \mbox{Im}\,\mathcal{T} \lvert B(p_B)\rangle\, ,
 \label{gammabuq1q2}
\end{eqnarray}
which we compute in the HQE up to $1/m_b^3$
\begin{equation}
\label{hqeTOPm3}
 \mbox{Im}\, \mathcal{T} = \Gamma^0_{\bar q_1 q_2} 
 \bigg( C_0 \mathcal{O}_0 
 + C_v \frac{\mathcal{O}_v}{m_b} 
 + C_\pi \frac{\mathcal{O}_\pi}{2m_b^2} 
 + C_G \frac{\mathcal{O}_G}{2m_b^2} 
 + C_D \frac{\mathcal{O}_D}{4m_b^3}
 + C_{LS} \frac{\mathcal{O}_{LS}}{4m_b^3}
 + \sum_{i,q} C_{4F_i}^{(q)} \frac{\mathcal{O}_{4F_i}^{(q)} }{4m_b^3}
 \bigg)\,,
\end{equation} 
where $p_B = M_B v$ is the $B$-hadron four-momentum, $m_b$ is the bottom quark 
mass, $\Gamma^0_{\bar q_1 q_2} = G_F^2 m_b^5 |V_{ub}|^2 |V_{q_1 q_2}|^2/(192\pi^3)$, $V_{ub}$ and $V_{q_1 q_2}$ are the corresponding CKM matrix elements, 
$q$ stands for a massless quark and
\begin{eqnarray}
\label{o0}
 \mathcal{O}_0 &=& \bar h_v h_v\,,
\\
 \mathcal{O}_v &=& \bar h_v (v\cdot \pi) h_v\,,
\\
 \mathcal{O}_\pi &=& \bar h_v \pi_\perp^2 h_v\,,
\\
 \mathcal{O}_G &=& \frac{1}{2}\bar h_v[\slashed \pi_{\perp}, \slashed \pi_{\perp}] h_v 
 = \frac{1}{2}\bar h_v [\gamma^\mu, \gamma^\nu] \pi_{\perp\,\mu}\pi_{\perp\,\nu}  h_v\,,
\\
 \mathcal{O}_D &=& \bar h_v[\pi_{\perp\,\mu},[\pi_{\perp}^\mu , v\cdot \pi]] h_v\,,
\\
\label{oLS}
 \mathcal{O}_{LS} &=& \frac{1}{2}\bar h_v[\gamma^\mu,\gamma^\nu]\{ \pi_{\perp\,\mu},[\pi_{\perp\,\nu}, v\cdot \pi] \} h_v\,,
\end{eqnarray}
are HQET local two-quark operators with $\pi_\mu= iD_\mu = i\partial_\mu +g_s A_\mu^a T^a$ and $\pi^\mu = v^\mu (v\pi) + \pi_\perp^\mu$. The four-quark operators 
$\mathcal{O}_{4F_i}^{(q)}$ will be defined in Sec. \ref{Sec:match4F}.

Note that the QCD spinor $b$ of the bottom quark is replaced by the HQET fermion field $h_v$. They are 
related as follows
\begin{equation}
\label{bhvtranf}
 b = e^{-im_b v\cdot x}\bigg[
 1 + \frac{\slashed \pi_\perp}{2m_b} 
 - \frac{(v\cdot \pi)\slashed \pi_\perp}{4m_b^2} 
 + \frac{\slashed\pi_\perp \slashed\pi_\perp}{8m_b^2} 
 + \frac{(v\cdot \pi)^2 \slashed \pi_\perp}{8m_b^3}
 + \frac{\slashed \pi_\perp \slashed\pi_\perp \slashed\pi_\perp}{16m_b^3}
 + \mathcal{O}(1/m_b^4)
 \bigg]h_v\,.
\end{equation}

\subsection{Matching of two-quark operators: computation of $C_i$}
\label{Sec:match2F}

\begin{figure}[!htb]  
	\centering
	\includegraphics[width=1.0\textwidth]{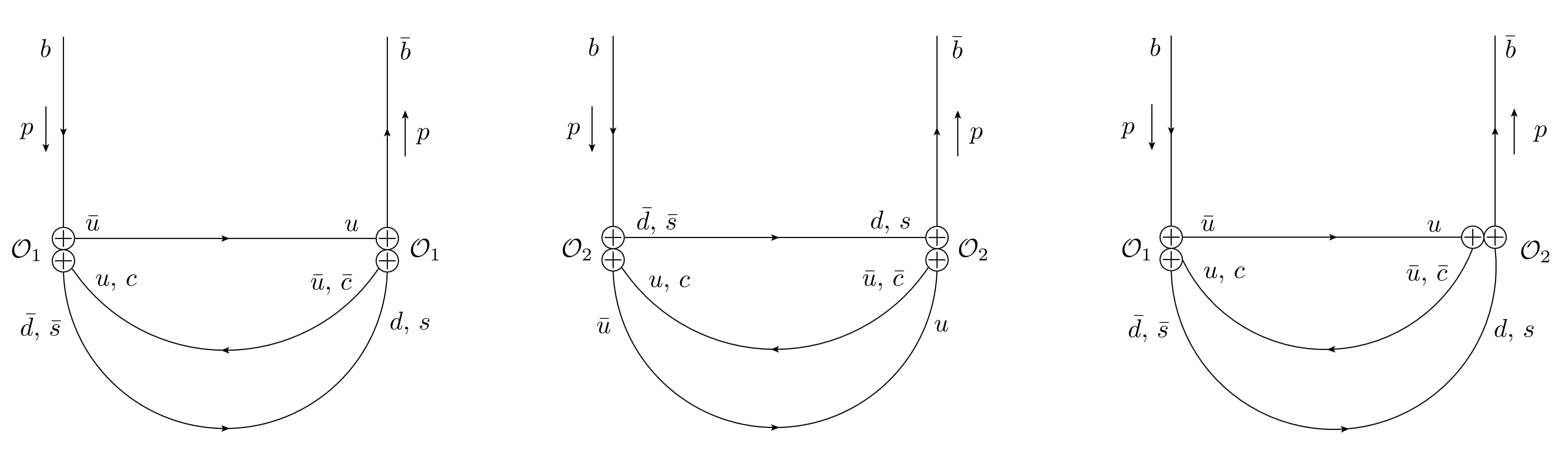}
\caption{Two-loop diagrams contributing to the matching coefficients of two-quark operators in the HQE of the 
$B$-hadron non-leptonic decay width.}
\label{forwardsme}   
\end{figure}

In this section we discuss the computation of the matching coefficients of Eq.~(\ref{hqeTOPm3}) associated to the two-quark 
operators Eqs.~(\ref{o0}-\ref{oLS}). To that purpose, we compute the $\mbox{Im}\,\mathcal{T}$ in the full theory 
and then match it with the HQE given in Eq.~(\ref{hqeTOPm3}). The Feynman diagrams representing the different contributions are 
shown in Fig.~[\ref{forwardsme}]. We take the $c$-quark to have mass $m_c$ and the $u,d,s$-quarks as massless. 
As a consequence, the matching coefficients will be just numbers for $b\rightarrow u\bar u d$ and will depend 
on the mass ratio $r=m_c^2/m_b^2$ for $b\rightarrow u \bar c s$. 

For the explicit calculation we find expressions for the quark propagator $S_F$ in an external gluon field \cite{NSVZ:84}, 
which automatically generates the proper ordering of the covariant derivatives, which is really convenient and optimal. 
We proceed as in Ref.~\cite{Mannel:2020fts} where a more detailed explanation is available, as well as explicit expressions for 
the quark propagators.

Therefore, the problem requires the computation of the imaginary part of two-loop diagrams of the sunset 
type with zero and one massive lines. We use LiteRed~\cite{Lee:2012cn,Lee:2013mka} to reduce integrals to a 
combination of a small set (indeed only one) of master integrals 
which we compute analytically (see Appendix). The Mathematica packages Tracer \cite{Jamin:1991dp} and HypExp \cite{Huber:2005yg,Huber:2007dx} are also used to deal 
with gamma matrices  and to perform the $\epsilon$ expansion of Hypergeometric functions, respectively.

Let's us discuss the peculiarities of each contribution:
\begin{itemize}
 \item {\bf $\mathcal{O}_1\otimes\mathcal{O}_1$ contribution}: The color structure only allows the radiation of a single gluon from 
 the $u$-quark in the ${\bar b} S_u b$ line. Therefore we only need to expand this particular $u$-quark propagator. 
 The computation is analogous to the semi-leptonic $b\rightarrow u \bar\ell\nu$ case. Due to the gluon emission from (or expansion of) a massless quark 
 propagator the coefficient of $\rho_D$ becomes IR divergent, which signals the mixing between the local operators of the HQE, in particular with the 
 four-quark operators. After renormalization, such IR divergences are canceled by UV divergences developed by the local operators in the effective theory.
 \item {\bf $\mathcal{O}_2\otimes\mathcal{O}_2$ contribution}: This case is identical to the previous one after the exchange of 
 the $u$ and $q_2$ quarks. Since both quarks are massless, the coefficients are invariant under the exchange of $C_1$ and $C_2$.
 \item {\bf $\mathcal{O}_1\otimes\mathcal{O}_2$ contribution}: It is precisely in this case where we observe the peculiarities of the 
 non-leptonic decay. Unlike in the previous cases, the gluon emission from (or expansion of) the quark propagators have to be taken into account from 
 all internal quark lines. Again, the gluon emission from (or expansion of) massless quark propagators will lead to IR divergences in the 
 coefficient of $\rho_D$ which properly cancel after renormalization. It is remarkable that in the $b\rightarrow u \bar u d$ case, where 
 all internal quark lines are massless, IR divergences cancel giving a finite coefficient. Finally, the computation of 
 $\mathcal{O}_1\otimes\mathcal{O}_2$ is found to be the same that for $\mathcal{O}_2\otimes\mathcal{O}_1$.
\end{itemize}

Overall, the coefficient of $\rho_{LS}$ is IR safe even for massless quarks, so it does not mix with the four 
quark operators. Note that it can be computed in four dimensions.

\subsection{Matching of four-quark operators: computation of $C_{4F_i}^{(q)}$}
\label{Sec:match4F}
\begin{figure}[!htb]  
	\centering
	\includegraphics[width=1.0\textwidth]{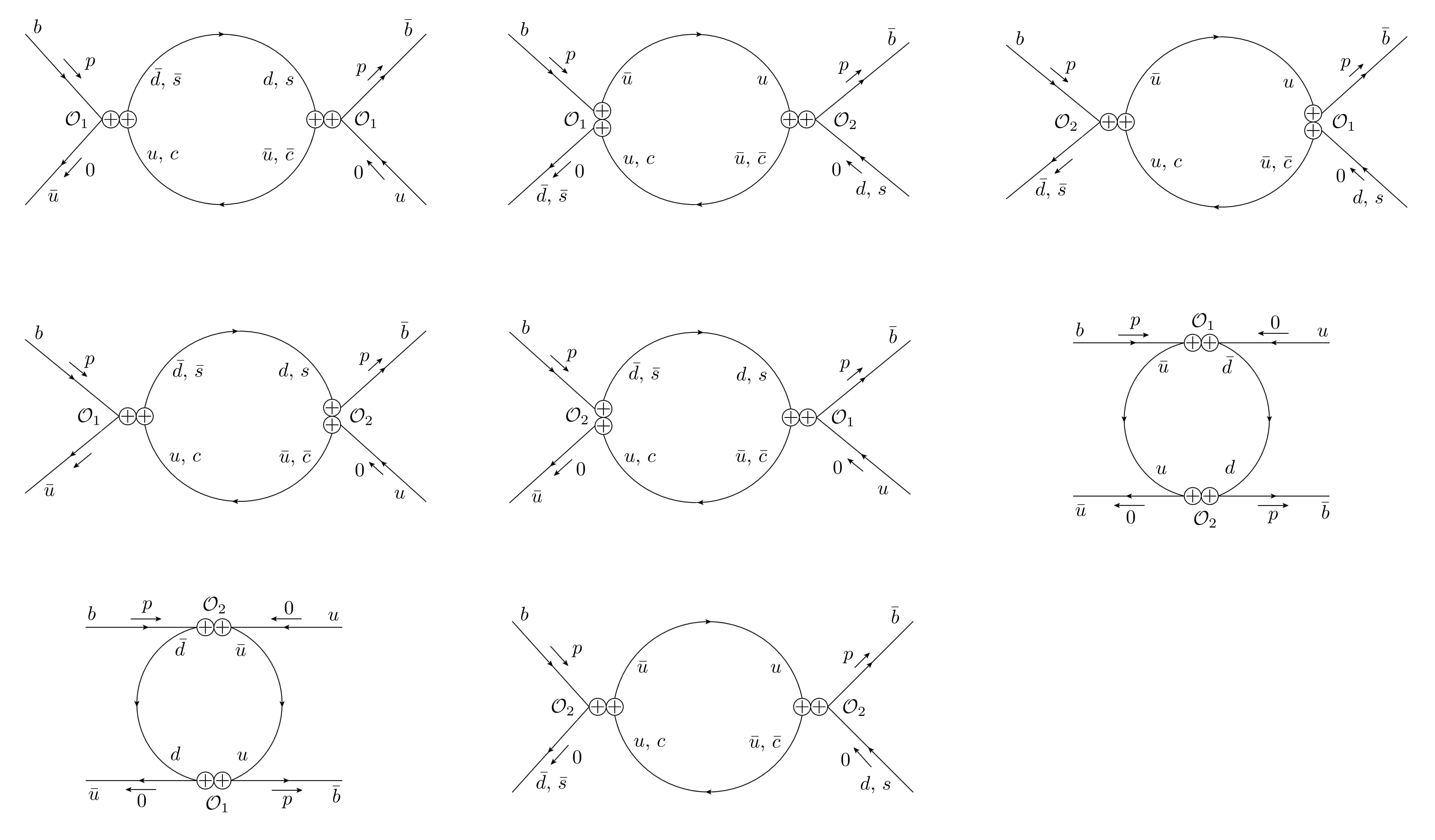}
\caption{One loop diagrams contributing to the tree level matching coefficients of four-quark operators in the HQE of the 
$B$-hadron non-leptonic decay width.}
\label{Matching4Fop}   
\end{figure}

In this section, we define the basis of four-quark operators $\mathcal{O}_{4F_i}^{(q)}$ appearing in Eq.~(\ref{hqeTOPm3}) which are 
relevant for the renormalization of $C_D$, 
and compute their matching coefficients at tree level. It requires the computation of the imaginary part of one-loop sunset type diagrams 
with zero and one massive lines. The diagrams contributing to the coefficients are shown in Fig.~[\ref{Matching4Fop}].

\subsubsection{The channel $b\rightarrow u\bar u d$}
\label{SubSec:match4qbcud}
The relevant operators in the HQE are
\begin{eqnarray}
 \mathcal{O}_{4F_1}^{(d)} &=& (\bar h_v \Gamma_\mu d) (\bar d \Gamma^\mu h_v)\,,
 \\
 \mathcal{O}_{4F_2}^{(d)} &=& (\bar h_v P_L d) (\bar d P_R h_v)\,,
 \\
 \mathcal{O}_{4F_1}^{(u)} &=& (\bar h_v \Gamma^\sigma \gamma^\mu \Gamma^\rho u) (\bar u \Gamma_\sigma \gamma_\mu \Gamma_\rho h_v)\,,
 \\
 \mathcal{O}_{4F_2}^{(u)} &=& (\bar h_v \Gamma^\sigma \slashed v \Gamma^\rho u) (\bar u \Gamma_\sigma \slashed v \Gamma_\rho h_v)\,,
 \\
 \tilde{\mathcal{O}}_{4F_1}^{(u)} &=& (\bar h_v \Gamma_\mu u) (\bar u \Gamma^\mu h_v)\,,
 \\
 \tilde{\mathcal{O}}_{4F_2}^{(u)} &=& (\bar h_v P_L u) (\bar u P_R h_v)\,,
\end{eqnarray}
where $P_L=(1-\gamma_5)/2$ and $P_R = (1+\gamma_5)/2$ are the left and right handed projectors, respectively. Their matching coefficients read  
\begin{eqnarray}
 C_{4F_1}^{(d)}  &=& -(3C_2^2 + 2C_1 C_2 (1-\epsilon) )\frac{3\cdot 2^{7+4\epsilon} \pi^{5/2+\epsilon} m_b^{-2\epsilon} (1-\epsilon)}{\Gamma(5/2-\epsilon)}
 \nonumber
 \\
 &=& -(3C_2^2 + 2C_1 C_2) 512\pi^2 \quad \mbox{for }\epsilon\rightarrow 0\,, 
 \\
 C_{4F_2}^{(d)} &=& (3C_2^2 + 2C_1 C_2(1-\epsilon) )\frac{3\cdot 2^{7+4\epsilon} \pi^{5/2+\epsilon} m_b^{-2\epsilon} (1-\epsilon)}{\Gamma(5/2-\epsilon)}
 \nonumber
 \\
 &=& (3C_2^2 + 2C_1 C_2) 512 \pi^2 \quad \mbox{for }\epsilon\rightarrow 0\,, 
 \\
 C_{4F_1}^{(u)} &=& C_1 C_2 \frac{3\cdot 2^{5+4\epsilon}  m_b^{-2\epsilon} \pi^{5/2+\epsilon}}{\Gamma(5/2-\epsilon)} 
 \nonumber
 \\
 &=& C_1 C_2 128 \pi^2 \quad \mbox{for }\epsilon\rightarrow 0\,, 
 \\
 C_{4F_2}^{(u)} &=& C_1 C_2 \frac{3\cdot 2^{6+4\epsilon}\pi^{5/2+\epsilon}  m_b^{-2\epsilon} (1-\epsilon)} {\Gamma(5/2-\epsilon)}
 \nonumber
 \\
 &=& C_1 C_2 256 \pi^2 \quad \mbox{for }\epsilon\rightarrow 0\,, 
 \\
 \tilde{C}_{4F_1}^{(u)} &=& -(3C_1^2 + 2C_1 C_2 (1-\epsilon) )\frac{3\cdot 2^{7+4\epsilon} \pi^{5/2+\epsilon} m_b^{-2\epsilon} (1-\epsilon)}{\Gamma(5/2-\epsilon)}
 \nonumber
 \\
 &=& -(3C_1^2 + 2C_1 C_2 ) 512\pi^2 \quad \mbox{for }\epsilon\rightarrow 0\,, 
 \\
 \tilde{C}_{4F_2}^{(u)} &=& (3C_1^2 + 2C_1 C_2(1-\epsilon) )\frac{3\cdot 2^{7+4\epsilon} \pi^{5/2+\epsilon} m_b^{-2\epsilon} (1-\epsilon)}{\Gamma(5/2-\epsilon)}
 \nonumber
 \\
 &=& (3C_1^2 + 2C_1 C_2 ) 512 \pi^2 \quad \mbox{for }\epsilon\rightarrow 0\,. 
\end{eqnarray}
These expressions coincide with the results of Ref.~\cite{Beneke:2002rj}.

\subsubsection{The channel $b\rightarrow u\bar c s$}

The relevant four-quark operators in the HQE are
\begin{eqnarray}
 \mathcal{O}_{4F_1}^{(s)} &=& (\bar h_v \Gamma_\mu s) (\bar s \Gamma^\mu h_v)\,,
 \\
 \mathcal{O}_{4F_2}^{(s)} &=& (\bar h_v P_L s) (\bar s P_R h_v)\,,
 \\
 \tilde{\mathcal{O}}_{4F_1}^{(u)} &=& (\bar h_v \Gamma_\mu u) (\bar u \Gamma^\mu h_v)\,,
 \\
 \tilde{\mathcal{O}}_{4F_2}^{(u)} &=& (\bar h_v P_L u) (\bar u P_R h_v)\,,
\end{eqnarray}
with matching coefficients

\begin{eqnarray}
 C_{4F_1}^{(s)} &=& -(3C_2^2 + 2C_1 C_2 (1-\epsilon) )\frac{3\cdot 2^{6+4\epsilon} \pi^{5/2+\epsilon} m_b^{-2\epsilon} (1-r)^{2-2\epsilon}(2+r-2\epsilon)}{\Gamma(5/2-\epsilon)}
 \nonumber
 \\
 &=& -(3C_2^2 + 2C_1 C_2) 256 \pi^2 (1-r)^2 (2+r) \quad \mbox{for }\epsilon\rightarrow 0\,, 
 \\
 C_{4F_2}^{(s)} &=& -(3C_2^2 + 2C_1 C_2(1-\epsilon))\frac{3\cdot 2^{7+4\epsilon} \pi^{5/2+\epsilon} m_b^{-2\epsilon} (1-r)^{2-2\epsilon} (-1+r(-2+\epsilon)+\epsilon)}{ \Gamma(5/2-\epsilon)}
 \nonumber
 \\
 &=& (3C_2^2 + 2C_1 C_2)512 \pi^2 (1-r)^2 (1+2r) \quad \mbox{for }\epsilon\rightarrow 0\,, 
 \\
 \tilde{C}_{4F_1}^{(u)} &=& -(3C_1^2 + 2C_1 C_2 (1-\epsilon) )\frac{3\cdot 2^{6+4\epsilon} \pi^{5/2+\epsilon} m_b^{-2\epsilon} (1-r)^{2-2\epsilon}(2+r-2\epsilon)}{\Gamma(5/2-\epsilon)}
 \nonumber
 \\
 &=& -(3C_1^2 + 2C_1 C_2) 256 \pi^2 (1-r)^2 (2+r) \quad \mbox{for }\epsilon\rightarrow 0\,,
 \\
 \tilde{C}_{4F_2}^{(u)} &=& -(3C_1^2 + 2C_1 C_2(1-\epsilon) )\frac{3\cdot 2^{7+4\epsilon} \pi^{5/2+\epsilon} m_b^{-2\epsilon} (1-r)^{2-2\epsilon} (-1+r(-2+\epsilon)+\epsilon)}{ \Gamma(5/2-\epsilon)}
 \nonumber
 \\
 &=& (3C_1^2 + 2C_1 C_2) 512 \pi^2 (1-r)^2 (1+2r) \quad \mbox{for }\epsilon\rightarrow 0\,,
\end{eqnarray}
Again, these expressions coincide with the results of Ref.~\cite{Beneke:2002rj}.

\subsection{Renormalization}
\label{Sec:renormalization}

The expression for the decay width given in Eq. (\ref{gammabuq1q2}) formally depends on the heavy quark mass $m_b$ and the 
QCD hadronization scale $\Lambda_{\rm QCD}$ with $m_b\gg \Lambda_{\rm QCD}$. The HQE given in Eq.~(\ref{hqeTOPm3}) takes advantage of the 
large scale separation and it is organized in such a
way that the coefficients are insensitive to the infrared scale $\Lambda_{\rm QCD}$, whereas the matrix elements of the local operators
are independent of the ultraviolet scale $m_b$.

However, a naive calculation produces both IR singularities in the coefficient functions and UV singularities
in the matrix elements of the local operators, which cancel at the end of the computation. We use dimensional regularization to deal with 
both IR and UV divergences. In such a setup, IR divergences in the coefficient functions
signal the UV mixing between the local operators of the HQE under renormalization.

\begin{figure}[!htb]  
	\centering
	\includegraphics[width=0.3\textwidth]{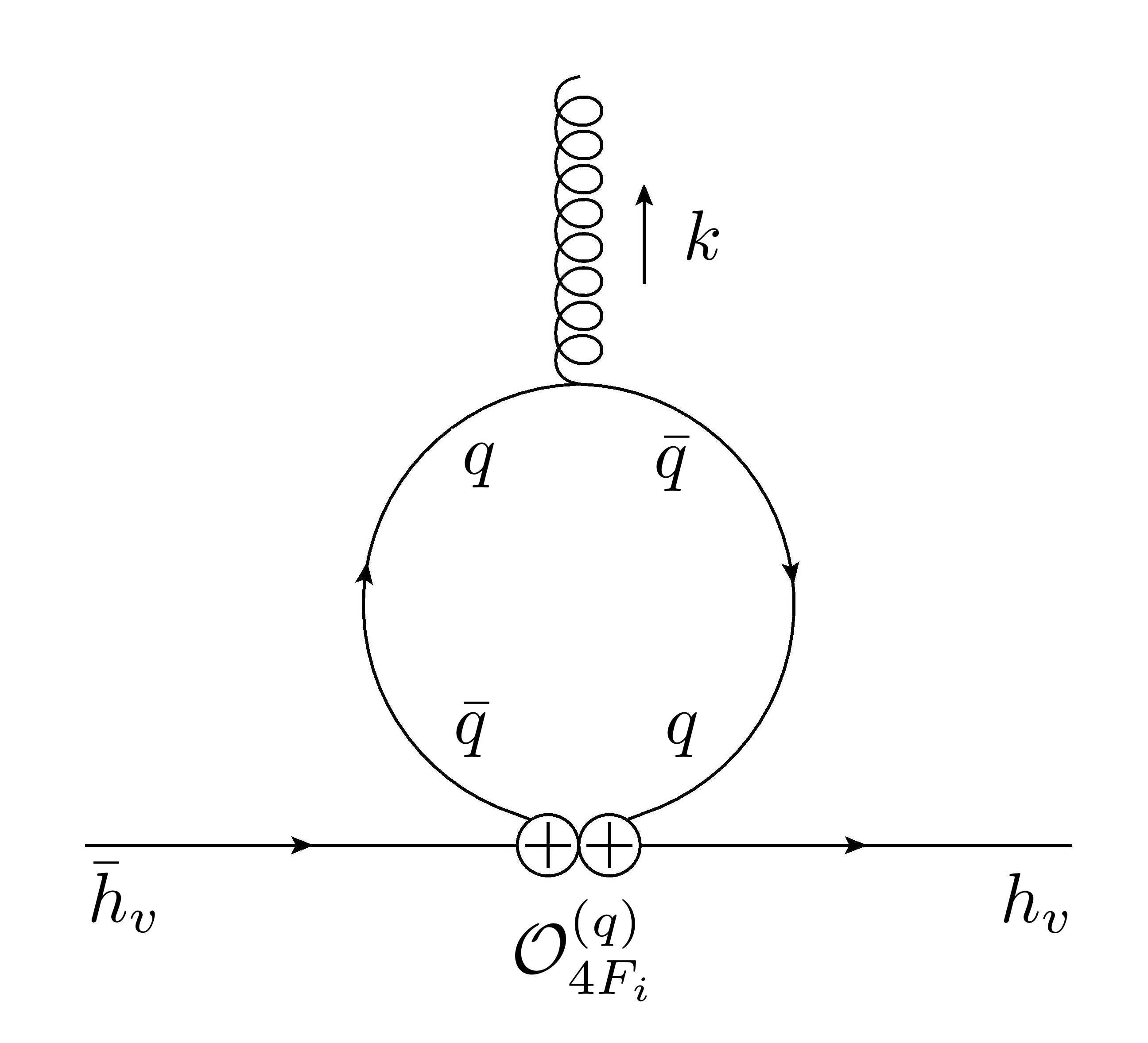}
\caption{One-loop diagrams contributing the renormalization of $C_{D}$.}
\label{ren}   
\end{figure}

In our computation a naive way of getting the coefficient of $\rho_D$ leads to an IR singularity due to the radiation of 
a soft gluon from masless quark lines. This singularity is canceled by the one-loop UV renormalization (see Fig. [\ref{ren}]) 
of the four-quark operators which has the general form (in the $\overline{\mbox{MS}}$ renormalization scheme)
\begin{equation}
\label{eq:gen4rhomixi}
 \mathcal{O}_{4F_i}^{(q)\,B}
=\mathcal{O}_{4F_i}^{(q)\,{\scriptsize\overline{\mbox{MS}}}}(\mu)
- \gamma_{4F_i}^{(q)}\frac{1}{48\pi^2\epsilon}\mu^{-2\epsilon}\bigg(\frac{e^{\gamma_E}}{4\pi}\bigg)^{-\epsilon} \mathcal{O}_D\,,
\end{equation}
where $\gamma_{4F_i}^{(q)}$ is the mixing anomalous dimension and $\bar\mu^{-2\epsilon} = \mu^{-2\epsilon}( e^{\gamma_E}/4\pi)^{-\epsilon}$ 
is the $\overline{\mbox{MS}}$ renormalization scale. In other words, 
The UV pole in $\epsilon$ coming from the operator mixing Eq.~(\ref{eq:gen4rhomixi}) between the four-quark operators and 
$\mathcal{O}_D$ cancels the IR divergence in the
coefficient function of $\rho_D$

\begin{equation}
 C_D^B = C_D^{{\scriptsize\overline{\mbox{MS}}}}(\mu)
 + \frac{1}{48\pi^2\epsilon}\mu^{-2\epsilon}\bigg(\frac{e^{\gamma_E}}{4\pi}\bigg)^{-\epsilon}\sum_{i,q} C_{4F_i}^{(q)} \gamma_{4F_i}^{(q)}\,.
\end{equation}
Let's define $\sum_{i,q} C_{4F_i}^{(q)} \gamma_{4F_i}^{(q)}\equiv {\bf C}_{4F_i}^{(q)} {\boldsymbol\gamma}_{4F_i}^{(q)}$.
Then for $b\rightarrow u\bar u d$ we obtain 

\begin{eqnarray}
 {\boldsymbol\gamma}_{4F_i}^{(q)} &=& (1,-1/2,16,4,1,-1/2)\,,
 \\
 {\bf C}_{4F_i}^{(q)} &=& (C_{4F_1}^{(d)},C_{4F_2}^{(d)},C_{4F_1}^{(u)},C_{4F_2}^{(u)},\tilde{C}_{4F_1}^{(u)},\tilde{C}_{4F_2}^{(u)})\,,
\end{eqnarray}
whereas for $b\rightarrow u \bar c s$ we obtain 

\begin{eqnarray}
 {\boldsymbol\gamma}_{4F_i}^{(q)} &=& (1,-1/2,1,-1/2)\,,
 \\
 {\bf C}_{4F_i}^{(q)} &=& (C_{4F_1}^{(s)},C_{4F_2}^{(s)},\tilde{C}_{4F_1}^{(u)},\tilde{C}_{4F_2}^{(u)})\,.
\end{eqnarray}
Likewise, we can write explicitely the counterterm of 
$C_D$ in the $\overline{\mbox{MS}}$ renormalization scheme. For $b\rightarrow u\bar u d$ it reads
\begin{equation}
 \delta C_D^{{\scriptsize\overline{\mbox{MS}}}}(\mu) = 
 \bigg( 
 C_{4F_1}^{(d)} - \frac{1}{2}C_{4F_2}^{(d)} 
 + 16C_{4F_1}^{(u)} + 4C_{4F_2}^{(u)} 
 + \tilde{C}_{4F_1}^{(u)} - \frac{1}{2}\tilde{C}_{4F_2}^{(u)}\bigg)
 \frac{1}{48\pi^2\epsilon}\mu^{-2\epsilon}\bigg(\frac{e^{\gamma_E}}{4\pi}\bigg)^{-\epsilon}\,,
\end{equation}
and for $b\rightarrow u \bar c s$ it reads

\begin{equation}
 \delta C_D^{{\scriptsize\overline{\mbox{MS}}}}(\mu) = 
 \bigg(
 C_{4F_1}^{(s)} - \frac{1}{2}C_{4F_2}^{(s)} 
+ \tilde{C}_{4F_1}^{(u)} - \frac{1}{2}\tilde{C}_{4F_2}^{(u)} 
 \bigg)
 \frac{1}{48\pi^2\epsilon}\mu^{-2\epsilon}\bigg(\frac{e^{\gamma_E}}{4\pi}\bigg)^{-\epsilon}\,,
\end{equation}
where $C_D^B = C_D^{{\scriptsize\overline{\mbox{MS}}}} + \delta C_D^{{\scriptsize\overline{\mbox{MS}}}}$.

\section{Results for the Wilson Coefficients up to order $1/m_b^3$}
\label{sect:results}

For the matching computation of the $\mbox{Im}\, \mathcal{T}$ we use the HQE given by Eq.~(\ref{hqeTOPm3}). 
However, for the final presentation of the results it is convenient to write it in terms of the 
local operator ${\bar b}\slashed v b$ defined in full QCD. Its HQE
\begin{equation}
\bar b \slashed v b = \mathcal{O}_0 - \tilde C_\pi \frac{\mathcal{O}_\pi}{2m_b^2} 
+ \tilde C_G \frac{\mathcal{O}_G}{2m_b^2} 
 + \tilde C_D \frac{\mathcal{O}_D}{4m_b^3}
 + \tilde C_{LS} \frac{\mathcal{O}_{LS} }{4m_b^3}
 + \mathcal{O}\left(\frac{\Lambda_{QCD}^4}{m_b^4} \right)\,.
\end{equation}
can be employed to remove the $\mathcal{O}_0$ operator which has the advantage that the forward matrix element of the leading term 
is normalized to all orders. We also use the equation of motion (EOM)
\begin{equation}
 \mathcal{O}_v =
 - \frac{1}{2m_b} (\mathcal{O}_\pi+ C_{{\scriptsize\mbox{mag}}}(\mu)\mathcal{O}_G)
 -  \frac{1}{8m_b^2} (c_D(\mu)\mathcal{O}_D + c_S(\mu)\mathcal{O}_{LS})
\end{equation}
to remove $\mathcal{O}_v$ from the expression for the $\mbox{Im}\,\mathcal{T}$. After these changes we obtain 
the following expression for the decay rate

\begin{eqnarray}
 \Gamma(b\rightarrow u \bar q_1 q_2) &=& \Gamma^0_{\bar q_1 q_2} 
 \bigg[ C_0 \bigg( 1 
- \frac{C_\pi  + C_0 \tilde C_\pi -  C_v}{C_0}\frac{\mu_\pi^2}{2m_b^2}\bigg)
+ \bigg(\frac{C_G - C_0 \tilde C_G}{C_{{\scriptsize\mbox{mag}}}(\mu)}-  C_v \bigg)\frac{\mu_G^2}{2m_b^2}
\nonumber
\\
&&
 - \bigg(\frac{C_D- C_0 \tilde C_D}{c_D(\mu)}-\frac{1}{2} C_v\bigg) \frac{\rho_D^3}{2m_b^3}
 - \bigg(\frac{C_{LS}- C_0 \tilde C_{LS} }{c_S(\mu)} - \frac{1}{2} C_v \bigg) \frac{\rho_{LS}^3}{2m_b^3}
\nonumber
\\
&&
 + \sum_{i,q} C_{4F_i}^{(q)} \frac{\langle\mathcal{O}_{4F_i}^{(q)}\rangle }{4m_b^3}
 \bigg]
\\
&\equiv& \Gamma^0_{\bar q_1 q_2} 
 \bigg[   C_0 
 - C_{\mu_\pi}\frac{\mu_\pi^2}{2m_b^2}
 + C_{\mu_G}\frac{\mu_G^2}{2m_b^2}
 - C_{\rho_D} \frac{\rho_D^3}{2m_b^3}
 - C_{\rho_{LS}} \frac{\rho_{LS}^3}{2m_b^3}
\nonumber
\\
&&
 + \sum_{i,q} C_{4F_i}^{(q)} \frac{\langle\mathcal{O}_{4F_i}^{(q)}\rangle }{4m_b^3}
 \bigg]\,.
\end{eqnarray}
in terms of the HQE parameters
\begin{eqnarray}
 \langle B(p_B)\lvert \bar b \slashed v b \lvert B(p_B)\rangle &=& 2M_B\,,  \\
 - \langle B(p_B)\lvert \mathcal{O}_\pi \lvert B(p_B)\rangle &=& 2M_B \mu_\pi^2\,, \\ 
 C_{{\scriptsize\mbox{mag}}}(\mu)\langle B(p_B)\lvert \mathcal{O}_G \lvert B(p_B)\rangle
 &=& 2M_B \mu_G^2\,, \\
  - c_D(\mu)\langle B(p_B)\lvert \mathcal{O}_D \lvert B(p_B)\rangle&=& 4M_B \rho_D^3\,, \\
 -  c_S(\mu)\langle B(p_B)\lvert \mathcal{O}_{LS} \lvert B(p_B)\rangle&=& 4 M_B \rho_{LS}^3\,, \\
 \langle B(p_B) \lvert\mathcal{O}_{4F_i}^{(q)}\lvert B(p_B)\rangle &=& 2M_B \langle\mathcal{O}_{4F_i}^{(q)}\rangle\,.
\end{eqnarray}
Note that $c_S=c_D=C_{{\scriptsize\mbox{mag}}}=1$ to the leading order. 
In the $\Gamma(b\rightarrow u\bar u d)$ case we find the following coefficients
\begin{eqnarray}
 C_0 = C_{\mu_\pi} &=& 3 C_1^2 + 2 C_1 C_2 + 3 C_2^2\,,
 \\
 C_v &=& 5 (3 C_1^2 + 2 C_1 C_2 + 3 C_2^2)\,,
 \\
 C_{\mu_G} = C_{\rho_{LS}} &=& -9 C_1^2 - 38 C_1 C_2 - 9 C_2^2\,,
 \\
 C_{\rho_D}^{{\scriptsize\overline{\mbox{MS}}}} &=&   
  -3 (C_1^2 + C_2^2) \bigg( 15 + 16 \ln\left(\frac{\mu^2}{m_b^2}\right) \bigg) 
  -14 C_1 C_2\,,
\end{eqnarray}
and for the case $\Gamma(b\rightarrow u\bar c s)$

\begin{eqnarray}
 C_0 = C_{\mu_\pi} &=& (3 C_1^2 + 2 C_1 C_2 + 3 C_2^2) (1 - 8 r + 8 r^3 - r^4 - 12 r^2 \ln(r) )\,,
 \\
 C_v &=& (3 C_1^2 + 2 C_1 C_2 + 3 C_2^2) (5 - 24 r + 24 r^2 - 8 r^3 + 3 r^4 - 12 r^2 \ln(r) )\,,
 \\
 C_{\mu_G} = C_{\rho_{LS}} &=& 
   (C_1^2 + C_2^2) (-9 + 24 r - 72 r^2 + 72 r^3 - 15 r^4 - 36 r^2 \ln(r) ) 
   \nonumber
   \\
   &&
 + 2 C_1 C_2 (-19 - 16 r + 24 r^2 + 16 r^3 - 5 r^4 - 12r(4 + r) \ln(r) ) \,,
 \\
 C_{\rho_D}^{{\scriptsize\overline{\mbox{MS}}}} &=& 
   (C_1^2 + C_2^2) \bigg(
     -45 + 16 r + 72 r^2 - 48 r^3 + 5 r^4 + 
     96 (-1 + r)^2 (1 + r) \ln(1 - r) 
\nonumber
     \\
     &&     
     + 12 (1 - 4 r) r^2 \ln(r) 
     - 48 (-1 + r)^2 (1 + r)\ln\left(\frac{\mu^2}{m_b^2}\right)
     \bigg)
\nonumber
     \\
     &&
     + \frac{2}{3} C_1 C_2 \bigg(
  107 - 152 r + 144 r^2 - 104 r^3 + 5 r^4 + 
     192 (-1 + r)^2 (1 + r) \ln(1 - r)
\nonumber
     \\
     &&
     + 12 (8 + 8 r + 5 r^2 - 8 r^3) \ln(r)
     - 96 (-1 + r)^2 (1 + r)\ln\left(\frac{\mu^2}{m_b^2}\right) \bigg)\,.
\end{eqnarray}
It is remarkable that by explicit calculation we obtain the relations $C_0 = C_{\mu_\pi}$ and $C_{\mu_G} =  C_{\rho_{LS}}$ 
which are a consequence of reparametrization 
invariance~\cite{Mannel:2018mqv}. This is a strong check of the calculation.

After using the transformation rules Eqs.~(3.20-3.23) in Ref.~\cite{Turczyk:07} we can compare our results with Ref.~\cite{Lenz:2020oce}, 
where the coefficients were computed in four dimensions. 
The coefficients up to $\mathcal{O}(1/m_b^2)$ are in agreement with Ref. \cite{Lenz:2020oce}, 
where coefficients are compared to the original papers. 
The coefficient of $\rho_D$ for $b\rightarrow u \bar c s$ is in agreement with Ref.~\cite{Lenz:2020oce}, as well. We postpone 
the comparison for $b\rightarrow u \bar u d$ to Sec.~\ref{Sec:eva}, as it requires a non-trivial change of the operator basis of four-quark operators 
which is related to the treatment of the Dirac algebra in $D$ dimensions \cite{Beneke:2002rj,Grozin:2017uto,Mannel:2020fts}.

\subsection{The canonical basis of four-quark operators}
\label{Sec:eva}
The results presented in Sec.~\ref{sect:results} for the $b \rightarrow u \bar u d$ channel are expressed in terms of the operators
\begin{eqnarray}
 \mathcal{O}_{4F_1}^{(u)} &=& 
(\bar h_v \Gamma^\sigma \gamma^\mu \Gamma^\rho u) (\bar u \Gamma_\sigma \gamma_\mu \Gamma_\rho h_v)
 = (\bar h_v \gamma^\sigma \gamma^\mu \gamma^\rho P_L u) (\bar u P_R \gamma_\sigma  \gamma_\mu \gamma_\rho h_v)\,,
\\ 
 \mathcal{O}_{4F_2}^{(u)} &=& 
(\bar h_v \Gamma^\sigma \slashed v \Gamma^\rho u) (\bar u \Gamma_\sigma \slashed v \Gamma_\rho h_v)\\
 &=& (\bar h_v \gamma^\sigma \gamma^\rho P_L u) (\bar u P_R \gamma_\sigma  \gamma_\rho h_v)
+ 4 (\bar h_v \gamma^\rho P_L u) (\bar u P_R \gamma_\rho h_v)
- 4 (\bar h_v P_L u) (\bar u P_R h_v)\,, \nonumber
\\
\tilde{\mathcal{O}}_{4F_1}^{(u)} &=& (\bar h_v \Gamma_\mu u) (\bar u \Gamma^\mu h_v)\,, 
\\
\tilde{\mathcal{O}}_{4F_2}^{(u)} &=& (\bar h_v P_L u) (\bar u P_R h_v)\,,
\end{eqnarray}
which constitute a basis in $D$-dimensions. However, one may want to use as a basis $\tilde{\mathcal{O}}_{4F_1}^{(u)}$ 
and $\tilde{\mathcal{O}}_{4F_2}^{(u)}$ only, like in Ref.~\cite{Lenz:2020oce}. 
While the two sets of operators can be related straightforwardly in $D = 4$, the situation for arbitrary $D$ is more 
involved. Doing so in $D$ dimensions requires the addition of new operators called evanescent operators, whose choice is not unique. 
A particular recipe reduces to the substitution~\cite{Beneke:2002rj,Grozin:2017uto}
\begin{eqnarray}
&&  \gamma_\mu\gamma_\nu\gamma_\alpha P_L \otimes \gamma^\mu \gamma^\nu \gamma^\alpha P_L \rightarrow 
 (16-a\epsilon)\gamma_\alpha P_L \otimes\gamma^\alpha P_L + E_1^{QCD}\,, \\ 
&&  \gamma_\mu\gamma_\nu P_L \otimes \gamma^\mu \gamma^\nu P_R \rightarrow 
 (4-b\epsilon)P_L \otimes P_R + E_2^{QCD}\,,
\end{eqnarray}
where $E_{1,2}^{QCD}$ are the so-called evanescent operators and $a$, $b$ are arbitrary numbers, which makes clear why 
the choice of the evanescent operators is not unique. A conventional choice is $a=4$ and $b=-4$, with $D=4-2\epsilon$. 
This choice is motivated by the requirement of validity of Fierz 
transformations at one-loop order~\cite{Buras:1989xd,Beneke:2002rj}. We will call the basis fixed by this choice 
to be the {\it canonical} basis of four-quark operators. The complete operator basis now reads
\begin{eqnarray}
\tilde{\mathcal{O}}_{4F_1}^{(u)} &=& (\bar h_v \Gamma_\mu u) (\bar u \Gamma^\mu h_v)\,, \\ 
\tilde{\mathcal{O}}_{4F_2}^{(u)} &=& (\bar h_v P_L u) (\bar u P_R h_v)\,, \\ 
 E_1^{QCD} &=& (\bar h_v \gamma_\mu\gamma_\nu\gamma_\alpha P_L u) (\bar u \gamma^\mu \gamma^\nu \gamma^\alpha P_L h_v)
 - (16-a\epsilon)(\bar h_v \Gamma_\mu u) (\bar u \Gamma^\mu h_v)\,, \\ 
 E_2^{QCD} &=& (\bar h_v \gamma_\mu\gamma_\nu P_L u) (\bar u P_R \gamma^\mu \gamma^\nu h_v) 
 - (4-b\epsilon)(\bar h_v P_L u)(\bar u P_R h_v)\,.
\end{eqnarray}
In the new basis the imaginary part of the transition operator becomes
\begin{equation}
 \mbox{Im}\, \mathcal{T}(b\rightarrow u\bar u d) = \Gamma^0_{\bar u d}
 \bigg(\ldots
 + \tilde{C}_{4F_1}^{(u)\,{\scriptsize\mbox{new}}} \frac{\tilde{\mathcal{O}}_{4F_1}^{(u)}}{4m_b^3}  
 + \tilde{C}_{4F_2}^{(u)\,{\scriptsize\mbox{new}}} \frac{\tilde{\mathcal{O}}_{4F_2}^{(u)}}{4m_b^3} 
 + C_{E_1}^{(u)} \frac{E_1^{QCD}}{4m_b^3}
 + C_{E_2}^{(u)} \frac{E_2^{QCD}}{4m_b^3}
 \bigg)\,,
\end{equation}
with the corresponding change in the counterterm of the coefficient $C_D$
\begin{equation}
 \delta C_D^{{\scriptsize\overline{\mbox{MS}}}} = 
 \bigg( 
 C_{4F_1}^{(d)} - \frac{1}{2}C_{4F_2}^{(d)} 
 + \tilde{C}_{4F_1}^{(u)\,{\scriptsize\mbox{new}}} - \frac{1}{2}\tilde{C}_{4F_2}^{(u)\,{\scriptsize\mbox{new}}} \bigg)
 \frac{1}{48\pi^2\epsilon}\mu^{-2\epsilon}\bigg(\frac{e^{\gamma_E}}{4\pi}\bigg)^{-\epsilon}\,,
\end{equation}
with
\begin{eqnarray}
\tilde{C}_{4F_1}^{(u)\,{\scriptsize\mbox{new}}} &=&  (16-a\epsilon) C_{4F_1}^{(u)} + 4C_{4F_2}^{(u)} + \tilde{C}_{4F_1}^{(u)} \,,
\\
 \tilde{C}_{4F_2}^{(u)\,{\scriptsize\mbox{new}}} &=& - b\epsilon C_{4F_2}^{(u)}  + \tilde{C}_{4F_2}^{(u)}\,,
\\
 C_{E_1}^{(u)} &=& C_{4F_1}^{(u)}\,,
\\
 C_{E_2}^{(u)} &=& C_{4F_2}^{(u)}\,.
\end{eqnarray}
The operators $E_{1,2}^{QCD}$ do not contribute to the anomalous dimension of 
$C_{D}$. The difference between the results obtained in the 
two bases is
\begin{equation}
 C_{\rho_D}^{{\scriptsize\overline{\mbox{MS}}}}(a,b) - C_{\rho_D}^{{\scriptsize\overline{\mbox{MS}}}}
 = \frac{8}{3}C_1 C_2 (a - b) \,,
\end{equation}
whereas the difference with the results presented in Ref.~\cite{Lenz:2020oce}, where the coefficient was computed in $D=4$, is
\begin{equation}
 C_{\rho_D}^{{\scriptsize\overline{\mbox{MS}}}}(a,b) - C_{\rho_D}^{{\scriptsize\overline{\mbox{MS}}},\,D=4} 
 = \frac{8}{3} C_1 C_2 (a - b - 8)  \,.
\end{equation}
Note that for the canonical choice of the evanescent operators the difference vanishes, so our results are in agreement with Ref.~\cite{Lenz:2020oce}.
As we mentioned there is some freedom when choosing the evanescent operators $E_{1,2}^{QCD}$. 
A different choice of $a$ and $b$ corresponds to the following shift in the coefficient 
\begin{equation}
 C_{\rho_D}^{{\scriptsize\overline{\mbox{MS}}}}(a_1,b_1) - C_{\rho_D}^{{\scriptsize\overline{\mbox{MS}}}}(a_2,b_2) 
 = \frac{8}{3} C_1 C_2 (a_1 - a_2 - b_1 + b_2)  \,.
\end{equation}

\subsection{Numerical evaluation}
\label{Sec:numerical}

In this section, we give numerical values for future phenomenological applications and as a first study of the expected size of 
the new contributions due to $\rho_D$ and $\rho_{LS}$. It is important to keep in mind that the $\rho_D$ coefficient depends on the scheme.
This scheme dependence is compensated by the matrix elements of the four-quark operators, which are also scheme-dependent. 
We use the $\overline{\rm MS}$ scheme for the definition of $\rho_D$ and chose for the scale $\mu = m_b$. 
The numbers given here must be considered as illustrative and a precise phenomenological analysis is postponed to future publications.
For the masses we use $m_b=4.8$ GeV and $m_c=1.3$ GeV. 

In tables~\ref{tabbuud} and~\ref{tabbuudcan} we give numerical values for the $b \rightarrow u\bar u d$ transition in 
the basis of Sec. ~\ref{SubSec:match4qbcud} and in the canonical basis, respectively. 
In table~\ref{tabbucs} we give numerical values for the $b \rightarrow u\bar c s$ transition. In the tables 
we show the coefficients in front of $C_1$ and $C_2$.
\begin{table}[!ht]
 \centering
  \begin{tabular}{|c|c|c|c|c|}
    \hline
    $b \rightarrow u\bar u d$ & $C_1^2$ & $C_2^2$ & $C_1 C_2$  \\ \hline
    $C_0$ & $3$ & $3$ & $2$ \\ \hline
    $C_{\mu_\pi}$ & $3$ & $3$ & $2$ \\ \hline
    $C_{\mu_G}$ & $-9$ & $-9$ & $-38$ \\ \hline
    $C_{\rho_D}$ & $-45$ & $-45$ & $-14$ \\ \hline
    $C_{\rho_{LS}}$ & $-9$ & $-9$ & $-38$ \\ \hline
    $C_{4F_1}^{(d)}/(128\pi^2)$ & $0$ & $- 12$ & $-8$ \\ \hline
    $C_{4F_2}^{(d)}/(128\pi^2)$ & $0$ & $12$ & $8$ \\ \hline
    $C_{4F_1}^{(u)}/(128\pi^2)$ & $0$ & $0$ & $1$ \\ \hline
    $C_{4F_2}^{(u)}/(128\pi^2)$ & $0$ & $0$ &  $2$ \\ \hline
    $\tilde{C}_{4F_1}^{(u)}/(128\pi^2)$ & $-12$ & $0$ & $- 8$ \\ \hline
    $\tilde{C}_{4F_2}^{(u)}/(128\pi^2)$ & $12$ & $0$ &  $8$ \\ \hline
    \end{tabular}
    \caption{Numerical values for the coefficients of $b \rightarrow u\bar u d$.}
    \label{tabbuud}
\end{table}
\begin{table}[!ht]
 \centering
  \begin{tabular}{|c|c|c|c|c|}
    \hline
    $b \rightarrow u\bar u d$ & $C_1^2$ & $C_2^2$ & $C_1 C_2$  \\ \hline
    $C_0$ & $3$ & $3$ & $2$ \\ \hline
    $C_{\mu_\pi}$ & $3$ & $3$ & $2$ \\ \hline
    $C_{\mu_G}$ & $-9$ & $- 9$ & $- 38$ \\ \hline
    $C_{\rho_D}$ & $-45$ & $-45$ & $22/3$ \\ \hline
    $C_{\rho_{LS}}$ & $-9$ & $- 9$ & $- 38$ \\ \hline
    $C_{4F_1}^{(d)}/(128\pi^2)$ & $0$ & $- 12$ & $-8$ \\ \hline
    $C_{4F_2}^{(d)}/(128\pi^2)$ & $0$ & $12$ & $8$  \\ \hline
    $\tilde{C}_{4F_1}^{(u)\,{\scriptsize\mbox{new}}}/(128\pi^2)$ & $-12$ & $0$ & $16$ \\ \hline
    $\tilde{C}_{4F_2}^{(u)\,{\scriptsize\mbox{new}}}/(128\pi^2)$ & $12$ & $0$ & $8$ \\ \hline
    $C_{E_1}^{(u)}/(128\pi^2)$ & $0$ & $0$ & $1$ \\ \hline
    $C_{E_2}^{(u)}/(128\pi^2)$ & $0$ & $0$ & $2$ \\ \hline
    \end{tabular}
    \caption{Numerical values for the coefficients of $b \rightarrow u\bar u d$ in the canonical basis ($a=4$, $b=-4$).}
    \label{tabbuudcan}
\end{table}
\begin{table}[!ht]
 \centering
  \begin{tabular}{|c|c|c|c|c|}
    \hline
    $b \rightarrow u\bar c s$ & $C_1^2$ & $C_2^2$ & $C_1 C_2$  \\ \hline
    $C_0$ & $1.75$  & $1.75$ & $1.17$ \\ \hline
    $C_{\mu_\pi}$ & $1.75$ & $1.75$ & $1.17$ \\ \hline
    $C_{\mu_G}$ & $-7.09$ & $- 7.09$ & $- 21.3$ \\ \hline
    $C_{\rho_D}$ & $-50.3$ & $- 50.3$ & $- 125$ \\ \hline
    $C_{\rho_{LS}}$ & $-7.09$ & $- 7.09$ & $- 21.3$ \\ \hline
    $C_{4F_1}^{(s)}/(128\pi^2)$ & $0$ & $- 10.7$ & $-7.12$ \\ \hline
    $C_{4F_2}^{(s)}/(128\pi^2)$ & $0$ & $11.8$ &  $7.88$ \\ \hline
    $C_{4F_1}^{(u)}/(128\pi^2)$ & $-10.7$  & $0$ & $- 7.12$ \\ \hline
    $C_{4F_2}^{(u)}/(128\pi^2)$ & $11.8$ & $0$ & $7.88$ \\ \hline
    \end{tabular}
    \caption{Numerical values for the coefficients of $b \rightarrow u\bar c s$.}
    \label{tabbucs}
\end{table}
For the Wilson coefficients of the weak hamiltonian we 
use the numerical values $C_1 (\mu_b) = -1.121$ and $C_2(\mu_b)  = 0.274$ at $\mu_b=4.0$ GeV given in Ref.~\cite{BBL:96}. 
We obtain

\begin{eqnarray}
 \frac{\Gamma_{\bar u d} }{\Gamma^0_{\bar u d}} &=&
 3.38
 - 1.69  \frac{\mu_\pi^2}{m_b^2} 
 - 0.14 \frac{\mu_G^2}{m_b^2} 
 + 27.8 \frac{\rho_D^3}{m_b^3} 
 + 0.14 \frac{\rho_{LS}^3}{m_b^3}
 + 492\frac{\langle\mathcal{O}_{4F_1}^{(d)}\rangle}{m_b^3}
 - 492\frac{\langle\mathcal{O}_{4F_2}^{(d)}\rangle}{m_b^3} 
 \nonumber
 \\
 &&
 - 97.4 \frac{\langle\mathcal{O}_{4F_1}^{(u)}\rangle}{m_b^3}
 - 195 \frac{\langle\mathcal{O}_{4F_2}^{(u)}\rangle}{m_b^3}
 - 3.98\cdot 10^3 \frac{\langle \tilde{\mathcal{O}}_{4F_1}^{(u)}\rangle}{m_b^3}
 + 3.98\cdot 10^3 \frac{\langle \tilde{\mathcal{O}}_{4F_2}^{(u)}\rangle}{m_b^3}\,,
\end{eqnarray}
\begin{eqnarray}
 \frac{\Gamma_{\bar u d}}{\Gamma^0_{\bar u d}}\bigg|_{{\scriptsize\mbox{can. basis}}} &=&
  3.38
 - 1.69 \frac{\mu_\pi^2}{m_b^2} 
 - 0.14 \frac{\mu_G^2}{m_b^2} 
 + 31.1 \frac{\rho_D^3}{m_b^3} 
 + 0.14 \frac{\rho_{LS}^3}{m_b^3}
 + 492\frac{\langle\mathcal{O}_{4F_1}^{(d)}\rangle}{m_b^3}
 - 492\frac{\langle\mathcal{O}_{4F_2}^{(d)}\rangle}{m_b^3}
 \nonumber
 \\
 && 
 - 6.32\cdot 10^3 \frac{\langle \tilde{\mathcal{O}}_{4F_1}^{(u)}\rangle}{m_b^3}
 + 3.98\cdot 10^3 \frac{\langle \tilde{\mathcal{O}}_{4F_2}^{(u)}\rangle}{m_b^3}
 - 97.4 \frac{\langle E_1^{QCD}\rangle}{m_b^3}
 - 195  \frac{\langle E_2^{QCD}\rangle}{m_b^3} \,,
\end{eqnarray}
\begin{eqnarray}
 \frac{\Gamma_{\bar c s} }{\Gamma^0_{\bar c s}} &=&
 1.98
 - 0.99\frac{\mu_\pi^2}{m_b^2} 
 - 1.44 \frac{\mu_G^2}{m_b^2} 
 + 14.3 \frac{\rho_D^3}{m_b^3} 
 +  1.44\frac{\rho_{LS}^3}{m_b^3}
 + 438 \frac{\langle\mathcal{O}_{4F_1}^{(s)}\rangle}{m_b^3}
 - 485\frac{\langle\mathcal{O}_{4F_2}^{(s)}\rangle}{m_b^3} 
  \nonumber
 \\
 &&
 - 3.55\cdot 10^3 \frac{\langle \tilde{\mathcal{O}}_{4F_1}^{(u)}\rangle}{m_b^3}
 + 3.92\cdot 10^3 \frac{\langle \tilde{\mathcal{O}}_{4F_2}^{(u)}\rangle}{m_b^3}\,,
\end{eqnarray}
where we used the abbreviation $\Gamma_{\bar q_1 q_2}$ to refer to $\Gamma(b\rightarrow u\bar q_1 q_2)$. Assuming that 
$\langle E_1^{QCD}\rangle/m_b^3=\langle E_2^{QCD}\rangle/m_b^3=0$ and 
$\rho_D^3/m_b^3 \sim \langle \mathcal{O}_{4F_i}^{(q)}\rangle/ m_b^3\sim \Lambda_{QCD}^3/m_b^3$, the Darwin coefficient gives a correction to the tree 
level values of the coefficients of the four-quark operators of $\sim 0.4\%$ for $b\rightarrow u\bar cs$ and of $\sim 0.5\%$ for $b\rightarrow u\bar ud$ 
in the canonical basis (we take the largest coefficient of the four-quark operators to compare). 
The coefficient of the spin-orbit term is less important since it is a factor ten smaller than the coefficient of the Darwin term. 
However, we have to keep in mind that the final size of the different terms is subject to the evaluation of the matrix elements, 
which is beyond the scope of this work. Finally, recall that not all four-quark operator contributions are included in the above expressions, but 
only those which were relevant for the renormalization of the Darwin term coefficient.

\section{Conclusions}
\label{sect:discussion}

In this paper, we have computed the coefficients of the $1/m_b^3$ hadronic matrix elements $\rho_D$ and 
$\rho_{LS}$ present in the HQE of the non-leptonic width of $B$-hadrons. 
More precisely, we have considered the Cabibbo suppressed contributions coming from $b\rightarrow u$ transitions.

The computation of the $\rho_{LS}$ coefficient is not specially interesting since reparametrization invariance 
fixes it to be equal to the $\mu_G$ coefficient, which is known. However, we compute it explicitly 
as a check.

The computation of the $\rho_D$ coefficient by itself is interesting from the theoretical point of view. 
It requires understanding the operator 
mixing between $\mathcal{O}_D$ and dimension six four-quark operators under renormalization.
As necessary ingredients, we have computed the leading order coefficients of the dimension six four-quark operators relevant 
to renormalize the $\rho_D$ coefficient. We find agreement with previously known results~\cite{Beneke:2002rj}.

We find that the $\rho_D$ coefficient depends on the calculational scheme. 
This not only concerns the use of the $\overline{\rm MS}$ scheme, but also the choice of the 
evanescent operators.

The coefficient of the Darwin term turns out to be sizable as it happened in the semi-leptonic and the non-leptonic ($b\rightarrow c$)
cases. It gives a correction to the tree 
level values of the coefficients of the four-quark operator of $\sim 0.4\%$ for $b\rightarrow u\bar c s$ and of $\sim 0.5\%$ for $b\rightarrow u\bar ud$ 
in the canonical basis. However, this statement requires some caveats. Since the coefficient is $\mu$-dependent we can only talk about its size 
after proper cancellation with the $\mu$-dependence of the matrix elements of the four-quark operators.
Indeed, the $\rho_D$ term by itself means nothing, but only its combination with the matrix elements of dimension six four-quark operators.
The coefficient of the spin-orbit term is less important since it is a factor ten smaller than the Darwin term coefficient.

The HQE has proven to be very successful to explain the lifetimes and lifetime differences of $B$-hadrons. 
Concurrently, experimental precision is becoming more impressive over the years. The results presented here may be relevant for a precise 
determination of $B$-hadron lifetimes and lifetime differences.

The results obtained here, specially for $b\rightarrow u\bar u d$, can also be easily applied to charm decays, where their contribution 
is expected to be more important due to the HQE for charm has a slower convergence than for bottom.
These coefficients might be helpful to clarify the status of the HQE for charm~\cite{Lenz:2013}, whose validity has been often questioned due to 
the smallness of the charm quark mass.

The $\rho_D$ term could be an important source of $SU(3)$ violation of lifetimes (lifetime 
differences)~\cite{Neubert:1996we,Mannel:2020fts}. That is because
$\rho_D$ can be related to the matrix elements of four-quark operators through the gluon EOM, 
whose matrix elements are not $SU(3)$ symmetric (depend on the spectator quark). 
However, a quantitative study of its impact on lifetime differences needs estimates 
of $SU(3)$ violation in the matrix elements, which is beyond the scope of this paper.

The coefficients computed in this work were recently determined in Ref.~\cite{Lenz:2020oce} using a different approach. After the proper choice
of the evanescent operators (canonical basis) we find agreement with the coefficients presented in this work.

\subsection*{Acknowledgments}
This research was supported by the Deutsche Forschungsgemeinschaft 
(DFG, German Research Foundation) under grant  396021762 - TRR 257 
``Particle Physics Phenomenology after the Higgs Discovery''. 
%
  
%
\section*{Technical results\label{sec:appendix}}
Here we collect some technical results used for the computations.

\subsection*{Master Integrals}

\subsubsection*{The decay $b \rightarrow u \bar c s$}

Let us define the two-loop integral
\begin{eqnarray}
 J(n_1,n_2,n_3,n_4,n_5) &\equiv& \int \frac{d^D q_1}{(2\pi)^D}\frac{d^D q_2}{(2\pi)^D}
 \frac{1}{(q_1^2)^{n_1} ((p_b+q_1-q_2)^2 - m_c^2)^{n_2} (q_2^2)^{n_3}((q_1 + p_b)^2)^{n_4}((q_2+p_b)^2)^{n_5}}
 \nonumber
 \\
 \label{basiceqIBP}
 &\equiv& \int \frac{d^D q_1}{(2\pi)^D}\frac{d^D q_2}{(2\pi)^D}\frac{1}{D_1^{n_1}D_2^{n_2}D_3^{n_3}D_4^{n_4}D_5^{n_5}}\,,
\end{eqnarray}
where $+i\eta$ prescriptions are assumed in the propagators. The following master integrals are needed for the computation
\begin{eqnarray} 
 J(0,1,0,1,1) &=& \int \frac{d^D q_1}{(2\pi)^D}\frac{d^D q_2}{(2\pi)^D}\frac{1}{((p_b+q_1-q_2)^2 - m_c^2) (q_1 + p_b)^2 (q_2+p_b)^2}
 \nonumber
 \\
 &=& \int \frac{d^D q_1}{(2\pi)^D}
  \frac{1}{((p_b-q_1)^2 - m_c^2)}
  \int \frac{d^d q_2}{(2\pi)^d}\frac{1}{(q_2 - q_1)^2 q_2^2}\,,
  \label{mi1sunset2loopMI}
 \\
 \label{mi2sunset2loopMI}
 J(0,2,0,1,1) &=& \int\frac{d^D q_1}{(2\pi)^D}\frac{d^D q_2}{(2\pi)^D}\frac{1}{((p_b+q_1-q_2)^2 - m_c^2)^{2} (q_1 + p_b)^2 (q_2+p_b)^2}\,.
\end{eqnarray}
Indeed, Eqs. (\ref{mi1sunset2loopMI}) and (\ref{mi2sunset2loopMI}) are not independent. Instead, they 
are related by a derivative with respect to $m_c^2$
\begin{equation}
 J(0,2,0,1,1) = \frac{d}{dm_c^2}J(0,1,0,1,1)\,.
\end{equation}
Therefore, we only need to compute explicitly one master integral, $J(0,1,0,1,1)$. More precisely, we need its imaginary part, which we 
denote by $\mbox{Im}\, J \equiv \bar J$. We obtain
\begin{equation}
 \bar J(0,1,0,1,1) = 
 \frac{2^{-8+4\epsilon} m_b^{2-4\epsilon}\pi^{-2+2\epsilon} (1-r)^{3-4\epsilon}\csc(\pi\epsilon)\Gamma(1-\epsilon)
 \,_2 F_1(2-2\epsilon,1-\epsilon;4-4\epsilon;1-r) }
 { \Gamma(\epsilon)\Gamma(4-4\epsilon)}\,,
\end{equation}
where $\,_2 F_1(a,b;c;z)$ is a hypergeometric function.

\subsubsection*{The decay $b\rightarrow u\bar u d$}

Let us define the two-loop integral
\begin{eqnarray}
 J(n_1,n_2,n_3,n_4,n_5)&\equiv& \int \frac{d^D q_1}{(2\pi)^D}\frac{d^D q_2}{(2\pi)^D}
 \frac{1}{(q_1^2)^{n_1} ((p_b+q_1-q_2)^2)^{n_2} (q_2^2)^{n_3}((q_1 + p_b)^2)^{n_4}((q_2+p_b)^2)^{n_5}}
 \nonumber
 \\
 \label{basiceqIBP}
 &\equiv& \int \frac{d^D q_1}{(2\pi)^D}\frac{d^D q_2}{(2\pi)^D}\frac{1}{D_1^{n_1}D_2^{n_2}D_3^{n_3}D_4^{n_4}D_5^{n_5}}\,,
\end{eqnarray}
where $+i\eta$ prescriptions are assumed in the propagators. The following master integral is needed
\begin{eqnarray} 
 J(0,1,0,1,1) &=& \int \frac{d^D q_1}{(2\pi)^D}\frac{d^D q_2}{(2\pi)^D}\frac{1}{(p_b+q_1-q_2)^2 (q_1 + p_b)^2 (q_2+p_b)^2}
 \nonumber
 \\
 &=& \int \frac{d^D q_1}{(2\pi)^D}
  \frac{1}{(p_b-q_1)^2}
  \int \frac{d^D q_2}{(2\pi)^D}\frac{1}{(q_2 - q_1)^2 q_2^2}\,,
  \label{mi1sunset2loopMIdos}
\end{eqnarray}
more precisely, its imaginary part. We obtain
\begin{equation}
 \bar J(0,1,0,1,1) = \frac{4^{-5 + 4\epsilon} m_b^{2 - 4 \epsilon} \pi^{-2 + 2 \epsilon}\Gamma(2 - 2 \epsilon) \Gamma(1 - \epsilon)}
  {\Gamma(3 - 3 \epsilon) \Gamma(3/2 - \epsilon)^2}\,.
\end{equation}

%

\newpage

\end{document}